\documentclass{emulateapj}
\usepackage{graphicx}
\usepackage{threeparttable}

\begin{document}
\newcommand\hst{\textit{HST }}
\newcommand\spitz{\textit{Spitzer }}
\newcommand\cf{$29.1^{+6.7}_{-5.3}\%$ }
\newcommand\cc{$11.0^{+7.0}_{-5.6}\%$ }
\newcommand\pc{$18.1\pm2.0\%$ }
\newcommand\ff{$35.2^{+5.2}_{-4.6}\%$ } 
\newcommand\fc{$24.7 ^{+5.3}_{-4.6}\%$ } 
\newcommand\pf{$10.5\pm 0.1\%$ }

\title{Galaxy Merger Candidates in High-Redshift Cluster Environments}

\author{A. G. Delahaye \altaffilmark{1},
T. M. A. Webb\altaffilmark{1},
J. Nantais\altaffilmark{2}
A. DeGroot\altaffilmark{3},
G. Wilson\altaffilmark{3},
A. Muzzin\altaffilmark{4},
H. K. C. Yee\altaffilmark{5},
R. Foltz\altaffilmark{3},
A. G. Noble\altaffilmark{6},
R. Demarco\altaffilmark{7},
A. Tudorica\altaffilmark{8},
M. C. Cooper\altaffilmark{9},
C. Lidman\altaffilmark{10},
S. Perlmutter\altaffilmark{11,12},
B. Hayden\altaffilmark{11,12},
K. Boone\altaffilmark{11,12},
J. Surace\altaffilmark{13}}

\altaffiltext{1}{Department of Physics, McGill University, 3600 Rue University, Montreal, QC, H3A 2T8, Canada}
\altaffiltext{2}{Departamento de Ciencias F\'{i}sicas, Universidad Andres Bello, Fernandez Concha 700, Las Condes 7591538, Santiago, Regin Metropolitana, Chile}
\altaffiltext{3}{Department of Physics \& Astronomy, University of California, Riverside, 900 University Avenue, Riverside, CA 92521, USA}
\altaffiltext{4}{Department of Physics and Astronomy, York University, 4700 Keele St., Toronto, Ontario, Canada, MJ3 1P3}
\altaffiltext{5}{Department of Astronomy \& Astrophysics, University of Toronto, 50 St. George Street, Toronto, ON, M5S 3H4, Canada}
\altaffiltext{6}{Kavli Institute for Astrophysics and Space Research, Massachusetts Institute of Technology, 77 Massachusetts Avenue, Cambridge, MA 02139, USA}
\altaffiltext{7}{Departamento de Astronom\'{i}a, Universidad de Concepc\'{i}on, Casilla 160-C, Concepc\'{i}on, Regin del Biobo, Chile}
\altaffiltext{8}{Argelander-Institute fur Astronomie, Auf dem Hugel 71, D-53121 Bonn, Germany}
\altaffiltext{9}{Center for Cosmology, Department of Physics and Astronomy, University of California, Irvine, 4129 Frederick Reines Hall, Irvine, CA 92697, USA}
\altaffiltext{10}{Australian Astronomical Observatory, PO Box 2915, North Ryde NSW 1670, Australia}
\altaffiltext{11}{Physics Division, Lawrence Berkeley National Laboratory, 1 Cyclotron Road, Berkeley, CA 94720, USA}
\altaffiltext{12}{Department of Physics, University of California Berkeley, 366 LeConte Hall MC 7300, Berkeley, CA 94720-7300, USA}
\altaffiltext{13}{Spitzer Science Center, California Institute of Technology, M/S 314-6, Pasadena, CA 91125, USA}

\email{delahaye@physics.mcgill.ca}

\begin{abstract}
We compile a sample of spectroscopically- and photometrically-selected cluster galaxies from four high-redshift galaxy clusters ($1.59 < z < 1.71$) from the \spitz Adaptation of the Red-Sequence Cluster Survey (SpARCS), and a comparison field sample selected from the UKIDSS Deep Survey. Using near-infrared imaging from the \textit{Hubble Space Telescope} we classify potential mergers involving massive ($M_* \geq 3\times 10^{10}\mathrm{M}_\odot$) cluster members by eye, based on morphological properties such as tidal distortions, double nuclei, and projected near neighbors within 20 kpc. With a catalogue of 23 spectroscopic and 32 photometric massive cluster members across the four clusters and 65 spectroscopic and 26 photometric comparable field galaxies, we find that after taking into account contamination from interlopers, \cc of the cluster members are involved in potential mergers, compared to \fc of the field galaxies. We see no evidence of merger enhancement in the central cluster environment with respect to the field, suggesting that galaxy-galaxy merging is not a stronger source of galaxy evolution in cluster environments compared to the field at these redshifts. 

\end{abstract}


\section{Introduction}
It is established in the local universe that the morphologies and star-formation rates of galaxies are related to both their intrinsic and extrinsic properties, such as mass and local environment \citep[e.g.,][]{kau03,kau04,pen10}. The mass of a galaxy has been shown to be correlated with colors, morphologies and specific star-formation rates (sSFRs) with more massive galaxies having lower sSFRs and older, redder, stellar populations \citep[for review]{balo04,bal06,bla09}. Additionally, high density clusters are populated by red-sequence ellipticals with low star-formation rates (SFR) and passively evolving stellar populations with a much earlier ($z\sim 2-4$) single epoch of active star-formation \citep{sta98,blak06,eis08,str10,ret11}. It is suggested that low-redshift cluster environments with extensive and hot intracluster gas and high densities cause gas depletion and thus quench star-formation in infalling galaxies \citep{bos06,bah15}. Thus, the majority of local universe star-formation occurs in lower density field populations which exhibit young stellar populations, bluer colors, and late-type morphologies \citep{dre80}. 

However, at some point at an earlier epoch stellar mass assembly must have occurred in cluster environments, likely coupled with increased star-formation. Studies of cluster environments at $z \sim 1$ have shown that the dependence of SFR on local density remains consistent with the paradigm seen in the local universe \citep{pat09,muz12}. Beyond a redshift of $z =1$, however, this relation weakens. Some studies have revealed a reversal in the SFR-density relation in non-cluster environments at these redshifts, most notably \citet{elb07,coo08}. Additionally, an increase in SFR with increasing redshift has been witnessed in the denser core regions of galaxy clusters \citep{tra10,bro13,san15}, even in the central brightest cluster galaxy regions \citep{web15b,mcd16}. 
Overdensities of submm sources at $z\sim 1-2$ suggest elevated levels of star formation over cluster and proto-cluster environments at high redshift \citep{cle14}.
Indeed, a transition epoch between unquenched and quenched SFRs at a redshift of $z\sim 1.4$ has been postulated by infrared studies of cluster populations \citep{man10,bro13,alb14}. SFRs in the cluster outskirts and into denser core regions are shown to be comparable or even enhanced relative to the field at these high redshifts, indicating a rapid quenching period occuring in clusters over $\sim z = 1.2 -1.4$ to explain the relatively quenched cluster populations we see by $z=1$ \citep{nan17}. The mechanisms responsible for this quenching are not thoroughly understood - possibilities include effects of the intracluster gas like ram-pressure stripping, intrinsic galaxy properties such as AGN feedback, or enhancement of galaxy-galaxy interactions including merging and harassment \citep{tre03}. 

Galaxy-galaxy mergers are favored in areas where there is an overdensity of galaxies and moderate relative velocities. If the relative velocities are too low, it will take too long (beyond a Hubble time) for coalescence to occur, and if the velocities are too high they will pass by each other, perhaps interacting but not able to merge \citep[see][for review]{mih04}. Galaxy clusters can provide high density environments where near neighbors are common - however, in the present day the velocity dispersions of massive virialized clusters are of the order 500-1000 km/s and not conducive to active merging amongst satellite galaxies. As would be expected, low redshift clusters are populated by mostly red and dead populations where star formation occurs only in the very outskirts or the field and merger rates in higher density regions are found to be on the order of 2-3\% \citep{ada12,cor16}. Any mergers occurring in cluster environments are likely dry mergers and do not contribute to the stellar mass assembly of the cluster via triggered star formation. Indeed, while dry merging may be evident in lower redshift clusters \citep{van99,tra05}, stellar mass assembly in clusters has been shown to be complete by moderate redshifts of at least $z > 1$ and possibly as distant as $z > 1.5$ \citep{and14b}, although it has been found that mass accretion of the brightest cluster galaxy in the central core is ongoing \citep[e.g.,][]{lid12}. 

Nonetheless, recent studies of high-redshift galaxy overdensities (termed `proto-clusters' by the authors) have seen evidence of enhanced merger rates, suggesting that merging may play an important role in mass assembly in these higher density environments. \citet{lot13} identified mergers in a $z=1.62$ proto-cluster and found an implied merger rate higher by a factor of 3-10 compared to the field. At even higher redshift, \citet{hin16} found elevated rates of merging Lyman-break galaxies in a $z=3.1$ proto-cluster, with a rate enhancement of over 60\% compared to the field. However, the merger rates in established clusters at high-redshift have not been investigated in detail, and it is unclear whether galaxy evolution in high-redshift cluster environments is dominated by local effects like active merging as suggested by \citet{man10}, or global effects like ram-pressure stripping or strangulation from the intracluster gas. 

In this paper we investigate the fraction of potential mergers in several high-redshift ($z > 1.5$) galaxy clusters, the largest study of its kind to date. We select four high-redshift galaxy clusters discovered in the \spitz Adaptation of the Red-Sequence Cluster Survey (SpARCS) cluster catalogue and spanning a redshift range $1.59 < z < 1.71$. All four clusters have been spectroscopically confirmed and have a wealth of multiwavelength observations, including deep near-infrared imaging from \textit{Hubble Space Telescope}. In \S 2 we summarize our datasets for both cluster and control, \S 3 outlines our merger identification method, our results are presented in \S 4 and discussed in \S 5. Throughout this paper we assume a $\Lambda$CDM cosmology with $H_0 = 70$km s$^{-1}$ Mpc$^{-1}$, $\Omega_M = 0.3$, and $\Omega_\Lambda = 0.7$.

\section{Data}
\subsection{Cluster Sample}
The push to identify galaxy clusters at these high-redshift epochs has resulted in the development of several novel observation techniques, and now dozens of galaxy clusters at redshifts $z> 1.3$ are known. The \spitz Adaptation of the Red-Sequence Cluster Survey \citep[SpARCS;][]{muz09,wil09} has provided over a dozen spectroscopically confirmed galaxy clusters at $z > 1.0$ and several at $z>1.5$. 
Our dataset comprises four rich galaxy clusters selected from the high-redshift cluster sample of the SpARCS catalogue, including selection based on the 1.6$\mu m$ Stellar Bump Sequence (SBS) method \citep{muz13}, that had been selected for extensive multiwavelength follow-up, including spectroscopy and \hst imaging.

SpARCS104922.6+564032.5 at $z=1.7089$ (hereafter J1049) was detected in the SpARCS coverage of the Lockman Hole and is notable for its highly star-forming brightest cluster galaxy \citep{web15a}. The remaining three clusters were all detected using SBS combined with SpARCS as described in \citet{muz13}. SpARCS033056--284300 ($z = 1.626$) and SpARCS022426-−032331 ($z=1.633$; hereafter J0330 and J0224) were both presented in \citet{lid12}, with J0224 additionally described in \citet{muz13}. SpARCS022546--035517  at $z=1.598$ (hereafter J0225) was presented in \citet{nan16}.  
The clusters are likely not fully virialized and the difficulty in obtaining spectroscopic redshifts for the passive members in the central core inhibits the ability to derive robust velocity dispersions; however, richness measurements suggest lower cluster mass limits of $10^{14}\mathrm{M}_\odot$. See Table \ref{clusters} for a summary of the cluster properties. 

\begin{table*}
\begin{center}
   \begin{threeparttable}
   \caption{Cluster Properties}
   \label{clusters}
   \begin{tabular}{l c c c c c c}
      \hline
      \hline
       & &  &  & Spec & \hst & Exposure \\
      Cluster ID & RA & Dec & $z$ & Members &Imaging & Times\\
      \hline
      SpARCS-J0225\tnote{a} & $02:25:45.6$ &$ -03:55:17.1$ & 1.598 & 8 & F160W & 2424s\\ 
      SpARCS-J0330\tnote{b} & $03:30:55.9$ & $-28:42:59.5$ & 1.626 & 38 & F105W, F140W, F160W & 10775s, 11625s, 5019s\\ 
      SpARCS-J0224\tnote{b,c} & $02:24:26.3$ & $-03:23:30.8$ & 1.633 & 45 & F105W, F140W, F160W & 7581s, 9829s, 6116s\\ 
      SpARCS-J1049\tnote{d}  & $10:49:22.6$ & $+56:40:32.6$ & 1.709 & 27 & F105W, F160W & 8543s, 9237s\\ 
      \hline

   \end{tabular}
   \begin{tablenotes}
      \item[a]{\citet{nan16}}
      \item[b]{\citet{lid12}}
      \item[c]{\citet{muz13}}
      \item[d]{\citet{web15a}}   
   \end{tablenotes}
   \end{threeparttable}
\end{center}
\end{table*}

Spectroscopic members were confirmed using the multi-object spectrometer MOSFIRE on Keck-I and the Focal Reduction and Imaging Spectrograph 2 (FORS2) on VLT \citep{muz13,web15a,nan16} with detailed reduction and analysis to be presented in DeGroot et al. (in prep). In total there are 118 confirmed spectroscopic members across the four clusters. 

Multiwavelength imaging is available from optical to infrared for all four clusters. 
For clusters J0224, J0225, and J0330, 11 to 12 band (not including additional \hst imaging described below) photometry is available. All three clusters have optical $ugriz$, near-infrared $YK_s$, and infrared (3.6$\mu$, 4.5$\mu$m, 5.8$\mu$m, 8.0$\mu$m), with additional near-infrared $J$ available for clusters J0224 and J0330\citep{nan16}. 
J1049 has 8 band photometry available with $ugrz$ from CFHT (Tudorica et al., in prep), $JK_s$ from \textit{UKIDSS} \citep{law07}, and IRAC 3.6$\mu$m, 4.5$\mu$m from SERVS \citep{mau12}. 
Photometric redshifts were determined using EAZY \citep{bra08} for all of the clusters using the above photometry with resulting normalized median absolute deviations ($\sigma_{nmad}$) of $(z_{phot}-z_{spec})/(1+z_{spec})$ of 0.04 for the 11/12 band photometry clusters \citep[J0224, J0225, J0330;][]{nan16} and 0.065 for the 8 band photometry cluster (J1049). Stellar masses were derived for all clusters using FAST \citep{kri09} with \citet{bru03} stellar population libraries, \citet{cal00} dust law, IMF from \citet{cha03}, and assuming an exponentially declining SFR. 

Deep \hst imaging was obtained for the central regions of the four clusters in the F160W filter on the WFC3-IR camera with additional imaging in F105W and F140W for a subset of the clusters  
from programs GO-14327, GO-13677 and GO-13747. 
Programs GO-13677 (cycle 22) and GO-14327 (cycle 23) were observed as part of the ``See Change'' program, a large \hst program 
using 174 orbits to discover and characterize $\sim$30 Type Ia supernovae at $z > 1$. The primary scientific goal of 
See Change is to improve our knowledge of the expansion history of the universe through distance 
measurements of high-redshift Type Ia supernovae, and calibration of the SZ-mass scaling relation 
using weak-lensing in the most massive, highest redshift clusters known to date.
For all four clusters, the \hst imaging covers a cluster-centric radius out to approximately 750 kpc. Standard reduction was performed on the images using the AstroDrizzle software available from the Space Telescope Science Institute. Reduced drizzled images have a pixel scale of 0.09" for all clusters with the exception of J0225 which has a final pixel scale of 0.128"/px. Exposure times in each filter are listed in Table \ref{clusters}. 

Cluster galaxies were selected based on both spectroscopic and photometric redshifts. Spectroscopic members required a spectroscopic redshift within 1000 km/s of their respective cluster redshift. For objects that had no spectroscopy, cluster members were selected based on high quality photometric redshifts (quality parameter $q_z <3$, as defined in \citet{bra08}). Photometric cluster members required the photometric redshifts to be within 2$\sigma_{nmad}$ of the cluster redshift, where $\sigma_{nmad}$ is 0.065 for J1049 and 0.04 for J0224, J0225, and J0330. Additionally, a mass cut was done requiring a stellar mass greater than $3\times 10^{10}M_\odot$ to ease comparison with other studies as well as ensuring completeness in all samples. The \hst imaging covers the central 750 kpc (in cluster-centric radius) region for each cluster and each galaxy was required to reside at least 3" (approximately 25 kpc physical) away from the edges of the \hst imaging to allow near neighbor analysis. The final mass-selected catalogue results in a total of 55 cluster members, 23 of those confirmed spectroscopically and 32 photometrically selected. 

\subsection{Control Sample}
To compare the fraction of merging systems to the field at high-redshift, we utilize the UKIDSS Deep Survey field (UDS), a pointing of the Cosmic Assembly Near-infrared Deep Extragalactic Survey (CANDELS)\citep{koe11}. Near-infrared \hst imaging from WFC3 is available for the UDS field in F125W and F160W to a two-orbits depth \citep{koe11,gro11,ske14}. The extensive spectroscopy \citep{bra12, mom15} provides a large sample of confirmed high-redshift ($z\sim 1.65$) galaxies. Complementary photometric analyses utilising a combination of ground-based and space-based observations provided 18 filter photometry to determine stellar masses \citep{ske14} using FAST \citep{kri09}, and derived with the same FAST parameters described above. 

We select massive ($M_* \geq 3\times 10^{10} M_\odot$) galaxies in the redshift range $1.55 < z < 1.75$ to sample the redshifts of our four clusters. Of those, 65 were selected via spectroscopic redshift and 26 selected via high quality ($q_z < 3$) photometric redshift. We do a regional cut to exclude the possibility of including members of the $z=1.62$ spectroscopically confirmed proto-cluster presented in \citet{pap10} and \citet{tra10}. Our final control catalogue consists of 91 galaxies.

\begin{figure*}
  \centering
    \includegraphics[width=\textwidth]{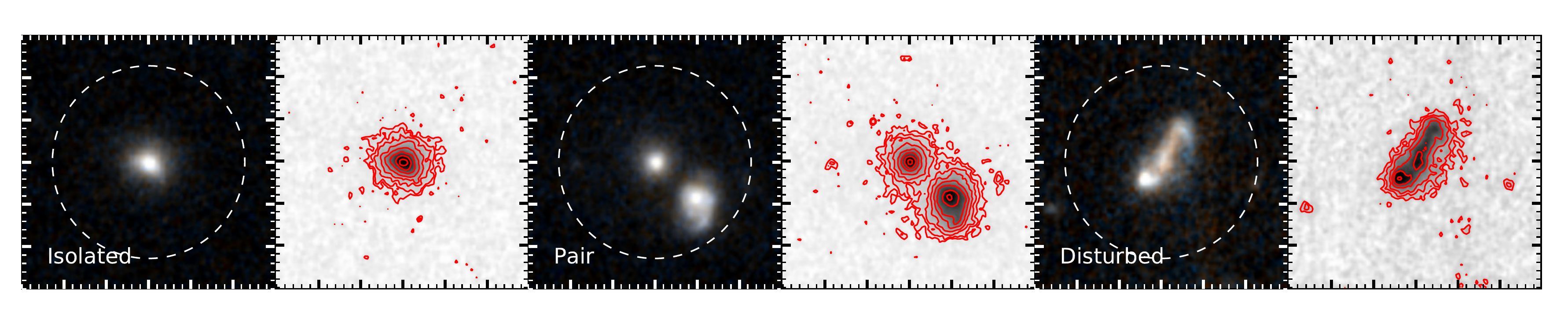}
   \caption{Examples of galaxies identified in each of the three classifications from the UDS control sample. Left panels are RGB images with filters F160W, F125W+F160W, F125W with a 20 kpc radius circle overlaid, and right panels are F160W grayscale maps with surface brightness shown as 0.5 mag arcsec$^{-2}$ contours. The galaxy in the left stamp is identified as isolated, with no near neighbor within the 20 kpc radius and no significant asymmetry or distortion. The galaxy in the central stamp has a clear near neighbor within 20 kpc. The galaxy in the right stamp shows signs of tidal distortion and strong asymmetry with no clear counterpart.}
   \label{classes}
\end{figure*}

\begin{figure*}
   \centering
      \includegraphics[width=\textwidth]{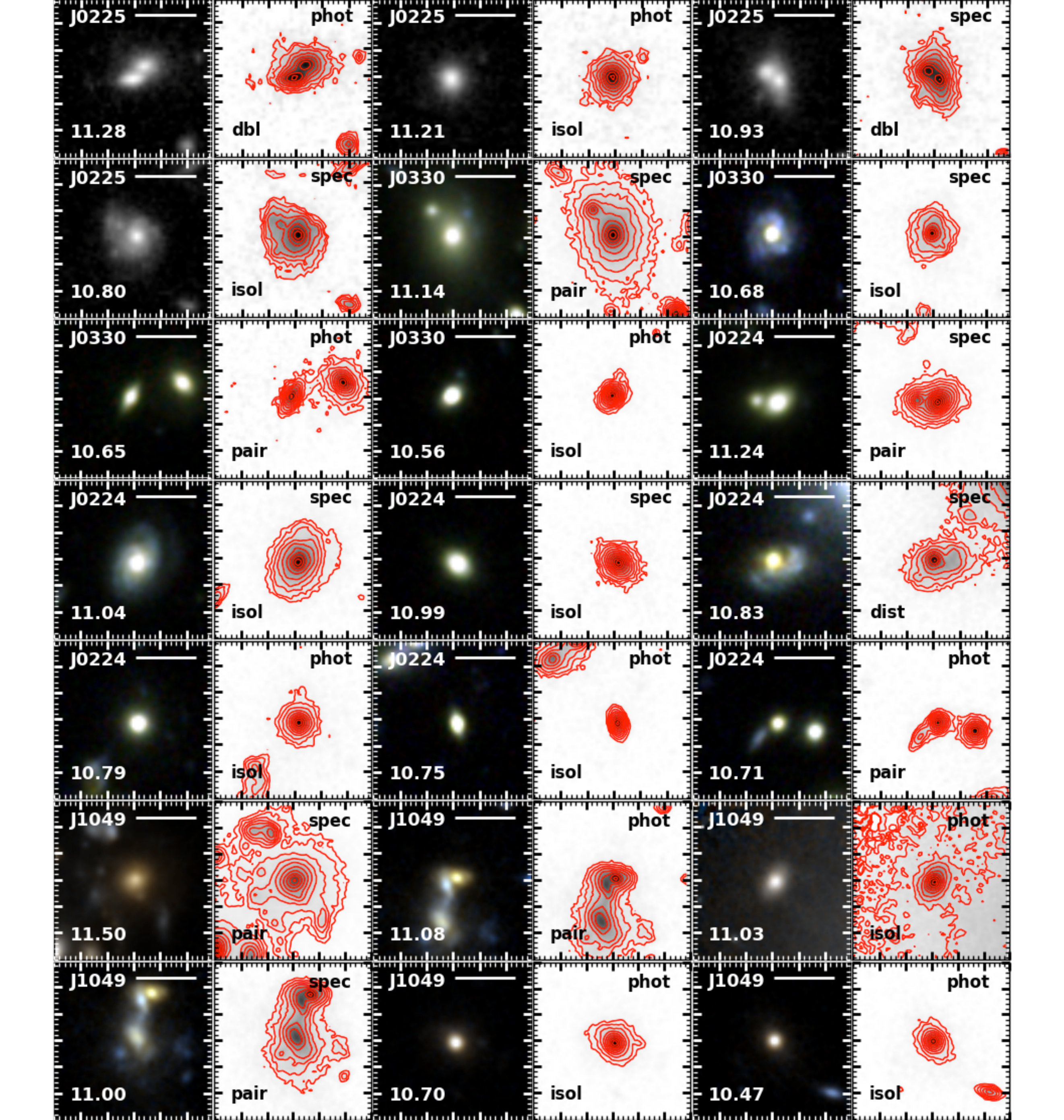}
   \caption{Simplified $6''\times 6''$ stamps of a subset of cluster members. The left panels show the cluster name in the top left, a horizontal bar indicating a distance of 20 kpc physical at the cluster redshift, and the log stellar mass in the lower left. The left images show grayscale F160W stamps for J0225, three-filter RGB (F105W, F140W, F160W) for clusters J0330 and J0224, and two-filter RGB in (F105W, F105W+F160W, F160W) for J1049. The right panels for all objects show grayscale F160W images with surface brightness contours starting at 24.5 mag arcsec$^{-2}$ and increasing by 0.5 mag arcsec$^{-2}$. The text in the upper right indicates whether the object was spectroscopically or photometrically selected, and the label in the lower left indicates the classification (isolated, pair, disturbed, or double nucleus).} 
   \label{stamps_cluster}
\end{figure*}

\begin{figure*}
   \centering
      \includegraphics[width=\textwidth]{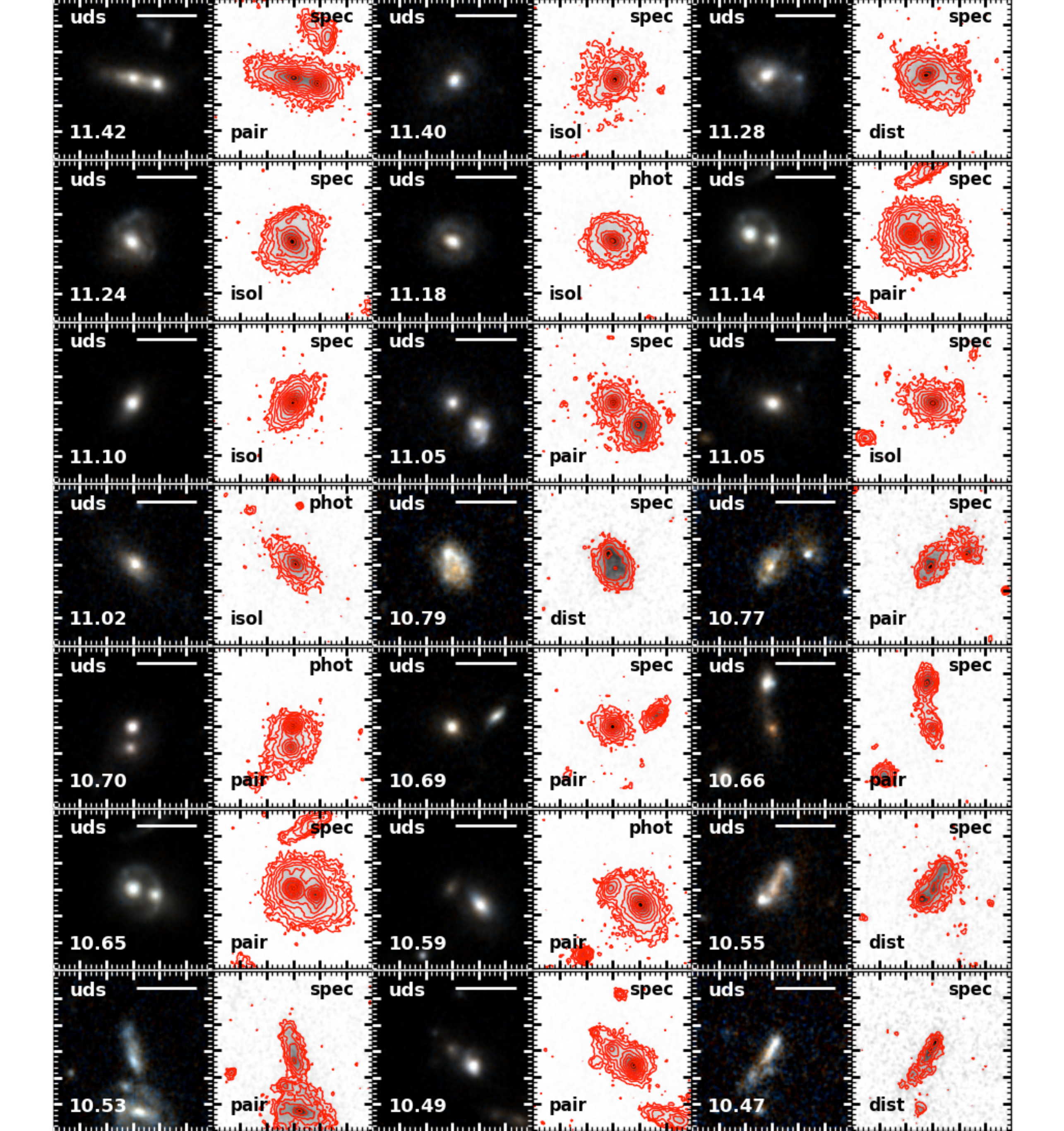}
   \caption{Same as Figure \ref{stamps_cluster} but for a subset of UDS galaxies. The left panels are two-filter RGB images (F160W, F125W+F160W, F125W) and the right show grayscale F160W stamps with same contours as Figure \ref{stamps_cluster}. Labels and text are the same as Figure \ref{stamps_cluster}.} 
   \label{stamps_uds}
\end{figure*}

\section{Merger Classification Methodology}
\hst imaging is available for all samples at a depth sufficient for morphological analyses: there is coverage in F160W for all clusters and the control sample in depths ranging from 2 to 9 orbits, as well as F125W for the control and some additional F105W and F140W for the cluster sample. The 4000\AA~ break spans from 1.02$\mu$m to 1.10$\mu$m across the redshift range $1.55 < z < 1.75$, so color images including the F105W band are likely highlighting different populations of stars compared to the redder filters and could introduce morphological differences not apparent in the F125W, F140W or F160W filters. Since F105W imaging is not available for all samples, for consistency the classifications were done on only single filter F160W maps. 

Each object was presented as a grayscale F160W $6''\times6''$ stamp with segmentation map contours from SExtractor overlaid. Two additional stamps displayed surface brightness contours in both finely (0.25 mag arcsec$^{-2}$) and broadly spaced contours (0.5 mag arcsec$^{-2}$), to highlight double nuclei, tidal features, and asymmetry. 
These subtle features can be difficult to identify and disentangle using automatic software and by-eye classification has been utilized in many classification surveys, taking advantage of the processing power and pattern recognition afforded by the human eye \citep[e.g.][]{lin08,kar15,will17}. Thus all objects were inspected and classified by eye to facilitate identification of features like close pairs, double nuclei, and asymmetries.
Identification of advanced mergers, such as the local universe post-merger Arp220, may be difficult if the nuclei are too close to distinguish or tidal features are too faint. However, the use of a control sample remedies this by looking for relative fractions of galaxies involved in mergers, rather than absolute numbers. To avoid bias during the classification process, each galaxy was inspected in randomized order so the location (field or cluster) was unknown during classification. J1049 has the deepest \hst imaging in F160W; so, to ensure that there were no biases towards faint features in the deep exposure, single orbit (900s) images of J1049 were also classified blindly. The classifications were consistent with one another regardless of depth; so, for our purposes the varying exposure times for different samples is not anticipated to be a significant issue. Classifications were done individually by three people - two team members and one non-team member. Classifications between all classifiers were consistent with one another and in ambiguous cases, the majority classification was used.

Galaxies were classified into three categories: Isolated, Pair, or Disturbed. Pairs were identified as having a near neighbor within a projected physical distance of 20 kpc, with no constraints on relative velocities, due to redshift incompleteness for the cluster sample. A magnitude limit of $m_{F160W} = 23.25$ for companions was used which roughly corresponds to a stellar mass of $10^{10} M_\odot$ at $z=1.65$ to select for systems likely to be major merger progenitors instead of minor merger progenitors. In cases where both pair members are present in the sample based on redshift and mass requirements, the system is counted twice. Disturbed galaxies have signs of merger activity, including tidal features such as tails or major asymmetry, or double nuclei present within the segmentation map. In many cases with double nuclei the secondary peak was not detected as a separate source so no magnitude requirement was placed on these systems. Isolated galaxies have no bright near neighbors or unusual morphology. Pairs and disturbed galaxies comprise our sample of `potential mergers'. For example images highlighting the different classifications, see Figure \ref{classes}. 


\begin{table*}
\begin{center}
   \begin{threeparttable}
   \caption{Merger Classification Results}
   \label{results}
   \begin{tabular}{l r r r r r r | r r r}
      \hline
      \hline\noalign{\smallskip}
      Sample & $N_{spec,m}$\tnote{a} & $N_{phot,m}$\tnote{a} & $N_{tot}$\tnote{b} & $N_{merg,s}$\tnote{c} & $N_{merg,p}$\tnote{c}& $N_{merg,tot}$\tnote{c}& $f_{merg}$ & $P_{int}$\tnote{d} & $f_{merg,c}$\tnote{e}\\
      \hline
      SpARCS-J0225 &  4    & 6  &   10 & 2    & 2 & 4    &-- &-- &-- \\ 
      SpARCS-J0330 &  3    & 6  &   9 & 1    & 1 & 2    &-- &-- &-- \\ 
      SpARCS-J0224 &  11   & 10 &   21 &  5   & 2  & 7    & -- &-- &-- \\ 
      SpARCS-J1049 &  5   & 10  &   15 &  2   & 1 &  3   & -- &-- &-- \\ 
      Cluster Total & 23  & 32  &  55  &  10  & 6  & 16 & \cf & \pc & \cc \\ [0.2cm]
      UDS Control &   65  & 26  &   91 &  24   & 7 &  31   & \ff & \pf & \fc \\[0.1cm] 
      \hline
   \end{tabular}
   \textbf{Notes}
   \begin{tablenotes}
      \item[a]{Number of objects spectroscopically or photometrically consistent with the cluster redshift, within \hst field-of-view, and with $M_* \geq 3\times 10^{10}\mathrm{M}_\odot$.}
      \item[b]{Total number of objects included in final sample for each field.}
      \item[c]{Number of galaxies in potential mergers, defined as either having a close pair with 20 kpc, or having signs of tidal features/double nuclei for spectroscopically or photometrically selected, and entire sample.}
      \item[d]{Probability of interlopers falsely contributing to projected pairs.}
      \item[e]{Fraction of galaxies in potential mergers after correcting for chance of interlopers.}
   \end{tablenotes}
   \end{threeparttable}   
\end{center}
\end{table*}

\begin{figure}
   \centering
      \includegraphics[width=0.5\textwidth]{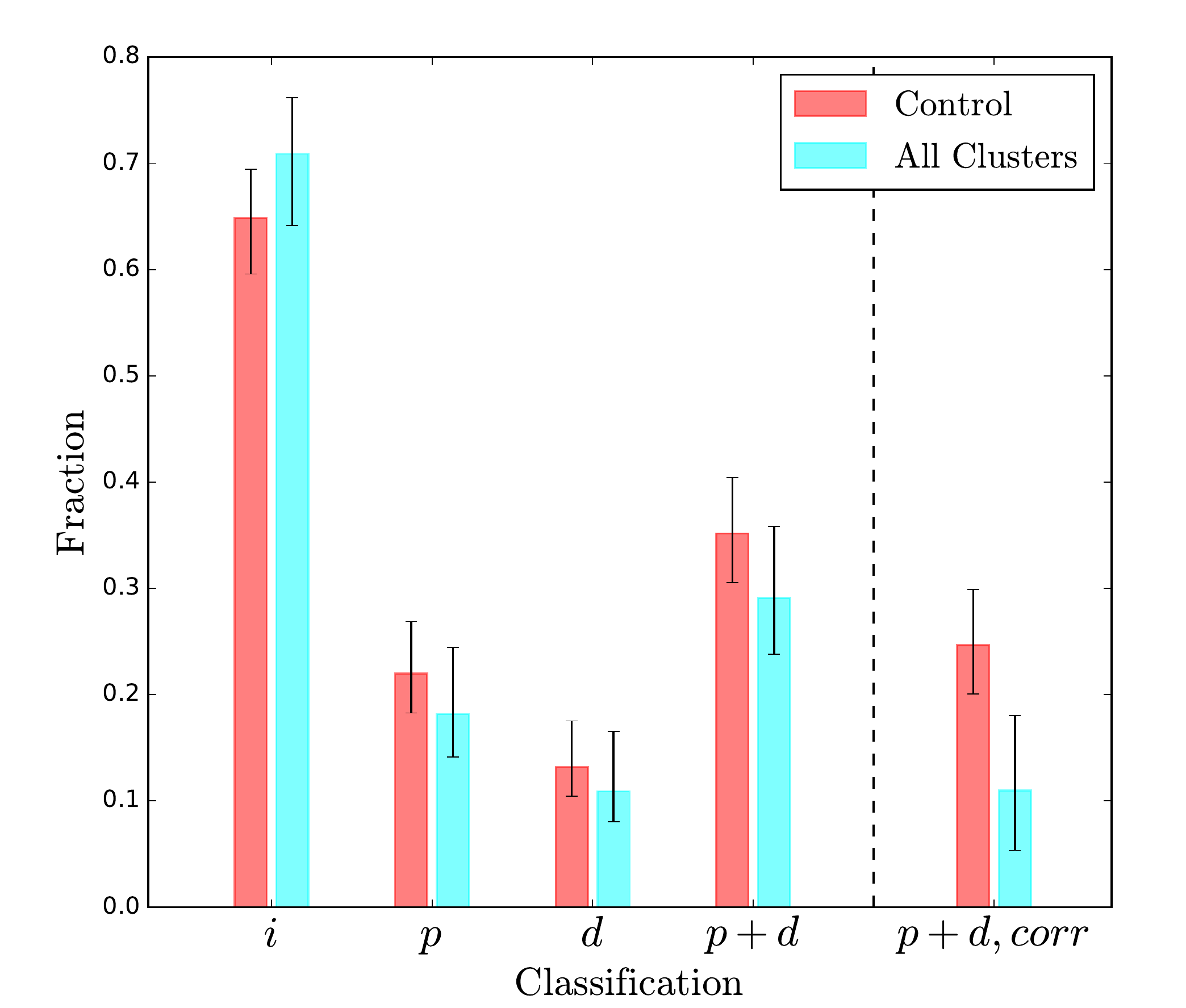}
      \caption{Fraction of galaxies found under each classification for UDS control sample and combined cluster sample. Left to right are isolated, pairs, disturbed, combined pairs + disturbed (representing overall merger probability) and combined pairs + disturbed after being corrected for interlopers. Errorbars show 68\% confidence and were estimated assuming a binomial distribution and utilizing the beta function.}
   \label{fracs}
\end{figure}

\section{Results}
Of the 55 redshift-selected cluster objects with $M_* \geq 3\times 10^{10} M_\odot$, 16 exhibit tidal features, double nuclei or close pairs, resulting in an observed fraction of merger candidates of \cf.
The UDS control sample comprising 91 objects with $M_* \geq 3\times 10^{10} M_\odot$, contains 31 objects with tidal features, double nuclei, or close pairs, resulting in an uncorrected fraction of merger candidates of \ff. Simplified versions of the stamps are shown for a subset of the cluster galaxies in Figure \ref{stamps_cluster}, and a subset of the control galaxies in Figure \ref{stamps_uds}.  Uncertainties were calculated assuming binomial statistics for 68\% confidence intervals, following \citet{cam11}. The results of the classification of both samples are presented in Figure \ref{fracs} and Table \ref{results}.

It is likely that some identified pairs are coincidental due to proximity along the line-of-sight and do not represent intrinsic nearby pairs. For the control sample of galaxies, this was corrected by randomly scattering all galaxies in the field with $m_{160} < 23.25$ and recording how often the nearest object for each of the massive, redshift-selected galaxies was within 20 kpc. This was iterated 1000 times resulting in an interloper fraction of \pf and a corrected merger candidate fraction of \fc. 

To preserve the overall distribution of galaxies within the clusters, where the higher densities can result in a higher probability of projected neighbors within the cluster but not necessarily interacting, the interloper fraction was calculated by randomly scattering all bright objects ($m_{160} < 23.25$) within a set radius of their current positions. The scattering radius was selected to be large enough that we are not just recovering the 20 kpc pairs and small enough that the scattering does not become completely random. A scattering radius of 11.5'' or 100 kpc physical at the clusters' redshift was chosen, although interloper fractions were calculated for scattering radii from 5'' to 20'' with the rate decreasing with increasing scattering radius and the lowest interloper fraction for uniform scattering. The combined interloper fraction across all four clusters is \pc resulting in a corrected merger candidate fraction of \cc.


\section{Discussion}
We have assembled a sample of 23 spectroscopically confirmed and 32 photometrically selected massive galaxies in four galaxy clusters spanning the redshift range $1.59 < z < 1.71$, along with a comparative control sample from UKIDSS Deep Survey comprising 65 spectroscopic and 26 photometric selected massive galaxies in the redshift range $1.55 < z < 1.75$. Through blind classification, we have identified 16 merger candidates in the cluster sample and 31 merger candidates in the control sample, resulting in potential merger fractions and 68\% confidence intervals of \cc and \fc for the cluster and control sample, respectively, after correcting for interlopers.
The potential merger fractions between the field and cluster samples are consistent within 1.6$\sigma$ although we cannot rule out the possibility that merger activity is suppressed in the core cluster environments by a factor of 2 or more. However, we can rule out the possibility of a mild enhancement of merger activity compared to field ($> 1.5$ times) at the $3\sigma$ level and a strong enhancement ($> 2$ times) at the $4\sigma$ level. 
 

Our sample is unique in that we are probing the environments of established clusters at high redshifts and our sample is significantly larger than similar studies. Previous studies by \citet{lot13} and \citet{hin16} have identified merger fractions in lower mass proto-cluster systems at $z = 1.62$ and $z=3.1$, finding elevated merger fractions by factors of roughly 5 and 1.5, respectively, when compared to the field. However, the enhancements are only significant at the $2\sigma$ and $1.5\sigma$ levels, respectively.
While all three studies find merger fractions between the field and cluster to be within $3\sigma$, we see no evidence of strongly elevated merger fractions in the clusters in contrast to the other two studies. 
A major difference between these studies and our own is the cluster environment - both \citet{lot13} and \citet{hin16} involve lower halo mass ($\sim 10^{13}\mathrm{M}_\odot$) proto-cluster environments, and we are probing the central regions (within 750 kpc) of massive, established clusters. Higher merger rates may be favored in proto-cluster environments where densities are higher than the field and infalling groups have low enough relative velocities to facilitate merging. 

The galaxy cluster halo mass has been suggested to play an important role in galaxy properties, with the halo mass dependence becoming stronger at higher redshifts. Simulations by \citet{mul15} indicate that with increasing redshift, the properties and halo distributions of current epoch massive clusters vary significantly. Variations in quenching efficiencies in cluster environments are found to be largest in higher redshift samples \citep{nan17}, suggesting halo mass or age may be dominant factors in galaxy evolution within cluster environments.
The halo mass may also directly play a role in the number of mergers seen \citep{bro13}, with more massive clusters assembling mass at earlier epochs whereas proto-clusters of the same epoch will still be assembling and accreting members. A larger halo mass will also deepen the gravitational potential well resulting in higher relative velocities in evolved systems which inhibit interactions and mergers between members. 
If cluster mass is indeed a driving factor in merger activity, it would be expected to see a higher merger fraction in lower mass halos at similar redshifts, which is evident in \citet{lot13} where the proto-cluster has a derived upper limit halo mass of several $10^{13}\mathrm{M}_\odot$ \citep{tan10,pie12}, an order of magnitude smaller than our cluster sample ($\log(M_*/\mathrm{M}_\odot) > 14.0$). 

As we see potential galaxy-galaxy merger fractions in central cluster regions comparable to the field, this suggests that merging is not a more dominant factor in the evolution of cluster galaxies relative to field populations. Our result is consistent with the conclusions drawn in \citet{and13} which suggest that mass build-up in massive cluster galaxies is mostly complete by $z\sim 1.8$ and enhanced build-up via merging in the redshift range $1.4 < z < 1.8$ is not expected in established clusters. The rapid quenching occuring in cluster populations from redshift $z\sim 1.6$ to $z\sim 1$ \citep[e.g.,][]{vdb13,dar16,nan16,nan17} is thus unlikely to be due to enhanced galaxy-galaxy merging, at least in the most massive cluster systems. This suggests that the driving forces in quenching cluster galaxies are more likely to be due to interactions within the intracluster medium (such as ram-pressure stripping), harassment, or mass-induced self-quenching \citep[e.g.,][]{pen10,bia15}. 

Simulations have shown that for all dark matter halos, regardless of mass, the overall halo merger rate (and implied galaxy merger rate) increases with increasing redshift \citep{wet09}. Yet the specific dependence on environment for merger rates is less studied. Overdense regions are found to have expected merger rates up to 2.5 times that of voids to a redshift of $z\sim 2$ \citep{fak09}, but whether that trend holds in the densest galaxy cluster cores is less certain. \citet{del07} have traced out the merger trees of brightest cluster galaxies (BCGs) and found that at large lookback times the BCG progenitor subhalos are undergoing mergers between themselves while in the cluster environment, before being eventually accreted onto the BCG itself. N-body simulations by \citet{ber09} have been used to trace back formation histories for dark matter halos in both cluster and field environments. They propose that at lookback times beyond 10 Gyr the merger rate of cluster subhalos may exceed the rate for merging field halos although this ratio drops significantly with decreasing redshift and most current cluster galaxies have not had a significant merger (greater than 1:10) within the past several Gyr. However, at $z=1.5$ the average rates for mergers become comparable between the two environments, consistent with the merger candidate fractions we determine in our $z\sim1.65$ cluster and control samples. 

We have derived the fraction of objects that appear to be undergoing or about to undergo mergers. The conversion from this fraction to an intrinsic merger rate is difficult, and depends on factors such as the relative velocities between pairs and timescales of tidal feature visibility. We have attempted to alleviate some of these complications by applying our requirements for identification equally across two samples (cluster and control) pulled from the same criteria in regards to stellar mass, redshift, and redshift selection method (both photometric and spectroscopic). As classification was done blindly between the two samples, any biases that could be introduced (for example due to individual classification by eye) should be present in both samples and thus not cause any discrepancies between the two. We have selected objects based on both spectroscopic and photometric redshift which allows identifications of both wet (star-forming, gas-rich) and dry (passive, quiescent) merging systems. In both spectroscopic and photometric samples the fraction of potential mergers is consistent with the corresponding field sample showing no enhancement for either wet or dry merging in these established clusters at $z\sim 1.65$. 

\acknowledgments
\noindent {\it Acknowledgments: }Based in part on observations obtained with MegaPrime/MegaCam, a joint project of CFHT and CEA/DAPNIA, at the Canada-France-Hawaii Telescope (CFHT), which is operated by the National Research Council (NRC) of Canada, the Institut National des Sciences de l’Univers of the Centre National de la Recherche Scientifique (CNRS) of France, and the University of Hawaii. This work is based in part on data products produced at TERAPIX and the Canadian Astronomy Data Centre as part of the Canada-France-Hawaii Telescope Legacy Survey, a collaborative project of NRC and CNRS. 

The author would like to thank the anonymous referee for their helpful comments and suggestions. The author also thanks J. Lowenthal for constructive discussions and assistance, as well as P. Scholz for aiding the analysis.

This work is based in part on observations taken by the 3D-HST Treasury Program (GO 12177 and 12328) with the NASA/ESA HST, which is operated by the Association of Universities for Research in Astronomy, Inc., under NASA contract NAS5-26555. TW acknowledges the support of an NSERC Discovery Grant. GW acknowledges financial support for this work from NSF grant AST-1517863 and from NASA through programs GO-13306, GO-13677, GO-13747 \& GO-13845/14327 from the Space Telescope Science Institute, which is operated by AURA, Inc., under NASA contract NAS 5-26555. JN acknowledges support from Universidad Andres Bello internal research project DI-18-17/RG. MCC acknowledges support for this work from NSF grant AST-1518257 and from NASA through grants GO-12547, AR-13242, and AR-14289 from the Space Telescope Science Institute, which is operated by AURA, Inc., under NASA contract NAS 5-26555. R.D. gratefully acknowledges the support provided by the BASAL Center for Astrophysics and Associated Technologies (CATA), and by FONDECYT grant No. 1130528.


\bibliographystyle{apj}
\bibliography{journals_apj,refs_mergers}

\end{document}